\documentclass[aip,apl,12pt]{revtex4-1}
\usepackage[version=3]{mhchem} % Formula subscripts using \ce{}
\usepackage[T1]{fontenc}       % Use modern font encodings
\usepackage[english]{babel}
\usepackage{graphicx}                      % immagini
\usepackage{amsmath,amssymb,amsthm}        % matematica
\usepackage[utf8]{inputenc}
\usepackage{textcomp}
\usepackage{amsmath}
\usepackage{color}

\newcommand{\tx}[1]{\textup{#1}}		% apici e pedici in text
\newcommand{\gtwo}{g^{(2)}(\tau)}
\newcommand{\gtwoZ}{g^{(2)}(0)}

\begin{document}

\title[]
  {Hybrid plasmonic waveguide coupling of photons from a single molecule}
%%%%%%%%%%%%%%%%%%%%%%%%%%%%%%%%%%%%%%%%%%%%%%%%%%%%%%%%%%%%%%%%%%%%%

\author{S.~Grandi}
\altaffiliation[Current address: ]{ICFO-Institut de Ciencies Fotoniques, The Barcelona Institute of Science and Technology, E-08860, Castelldefels (Barcelona), Spain\\}
\affiliation{Centre for Cold Matter, Blackett Laboratory, Prince Consort Road, South Kensington, SW7 2AZ, UK}

\author{M.~P.~Nielsen}
\altaffiliation[Current address: ]{School of Photovoltaic and Renewable Energy Engineering, University of New South Wales, Sydney, NSW, 2052, Australia\\}
\affiliation{Experimental Solid State Group, Blackett Laboratory, Prince Consort Road, South Kensington, SW7 2AZ, UK}

\author{J. Cambiasso}
\affiliation{Experimental Solid State Group, Blackett Laboratory, Prince Consort Road, South Kensington, SW7 2AZ, UK}

\author{S.~Boissier}
\affiliation{Centre for Cold Matter, Blackett Laboratory, Prince Consort Road, South Kensington, SW7 2AZ, UK}

\author{K.~D.~Major}
\affiliation{Centre for Cold Matter, Blackett Laboratory, Prince Consort Road, South Kensington, SW7 2AZ, UK}

\author{C.~Reardon}
\affiliation{University of York, York, UK}

\author{T.~F.~Krauss}
\affiliation{University of York, York, UK}

\author{R.~F.~Oulton}
\affiliation{Experimental Solid State Group, Blackett Laboratory, Prince Consort Road, South Kensington, SW7 2AZ, UK}

\author{E.~A.~Hinds}
\affiliation{Centre for Cold Matter, Blackett Laboratory, Prince Consort Road, South Kensington, SW7 2AZ, UK}

\author{A.~S.~Clark}
\email[email: ]{alex.clark@imperial.ac.uk}
\affiliation{Centre for Cold Matter, Blackett Laboratory, Prince Consort Road, South Kensington, SW7 2AZ, UK}

%%%%%%%%%%%%%%%%%%%%%%%%%%%%%%%%%%%%%%%%%%%%%%%%%%%%%%%%%%%%%%%%%%%%%
\date{\today}

\begin{abstract}
We demonstrate the emission of photons from a single molecule into a hybrid gap plasmon waveguide (HGPW). Crystals of anthracene, doped with dibenzoterrylene (DBT), are grown on top of the waveguides. We investigate a single DBT molecule coupled to the plasmonic region of one of the guides, and determine its in-plane orientation, excited state lifetime and saturation intensity. The molecule emits light into the guide, which is remotely out-coupled by a grating. The second-order auto-correlation and cross-correlation functions show that the emitter is a single molecule and that the light emerging from the grating comes from that molecule. The coupling efficiency is found to be $\beta_{WG}=11.6(1.5)\%$. This type of structure is promising for building new functionality into quantum-photonic circuits, where localised regions of strong emitter-guide coupling can be interconnected by low-loss dielectric guides.

\end{abstract}


\maketitle %\maketitle must follow title, authors, abstract and \pacs

%%%%%%SECTION
\section{Introduction}
Despite great advances over the last decade, the wider uptake of quantum technology has been inhibited by the lack of an efficient single photon source. Among several candidates\cite{Aharonovich2016}, single molecules are promising as a way to deliver narrow-band photons rapidly and on demand\cite{Nicolet2007a,Trebbia2009,Faez2014,Wang2017}. A variety of molecules are photostable and several wavelengths are available by choosing suitable combinations of dopant and host\cite{Tamarat2000}. While fulfilling most of the requirements for quantum technologies,\cite{Eisaman2011} molecules naturally emit light into a range of directions, as do most emitters\cite{Aharonovich2016}, and therefore collection of the photons requires some attention.  The use of micro-pillars has been a very successful approach\cite{Hausmann2011,Ding2016,Somaschi2016}, but is not naturally suited to building optical circuits as the photons are extracted perpendicular to the chip. Dielectric waveguides can encourage emission into the plane of the chip\cite{Hwang2011,Faez2014,Polisseni2016,Turschmann2017} but good coupling requires the emitters to be placed inside the guide\cite{Xu2004,Robinson2005,Barrios2007}, which is a challenge, and even then the coupling is limited by the transverse mode area of the guide. Plasmonic waveguides can have much smaller mode areas, but are compromised by absorption losses and non-radiative decay of the emitter\cite{Dieleman2017,Siampour2017}. Plasmonic antennas can help by concentrating the field at the site of the emitter into a much smaller volume, and can redirect the emission into a well-controlled direction. Indeed, this idea was demonstrated with a Yagi-Uda antenna\cite{Lee2011,Checcucci2017}, but did not direct the light into a single optical mode and, like the micropillar, is not naturally compatible with a planar integrated architecture.
Here we have taken a hybrid dielectric--metal approach\cite{Lafone2014}, using a planar hybrid gap plasmon waveguide (HGPW). The propagating hybrid optical mode switches from mostly dielectric to mostly plasmonic and back to mostly dielectric, coupling to a single molecule of dibenzoterrylene (DBT) in the plasmonic region. There, the small transverse mode area enhances photon emission into the waveguide, after which the photon moves into the low-loss dielectric region. This structure can provide both high coupling and low loss, making it suitable for an optical network, where single emitters interact with the waveguide field in selected ``hot spots'', while low-loss propagation interconnects them.
%%%%%%SECTION
\section{HGPW design, fabrication, and functionalisation}
%%%%%%FIGURE
\begin{figure*}[t!]
	\centering
	\includegraphics[width=\columnwidth]{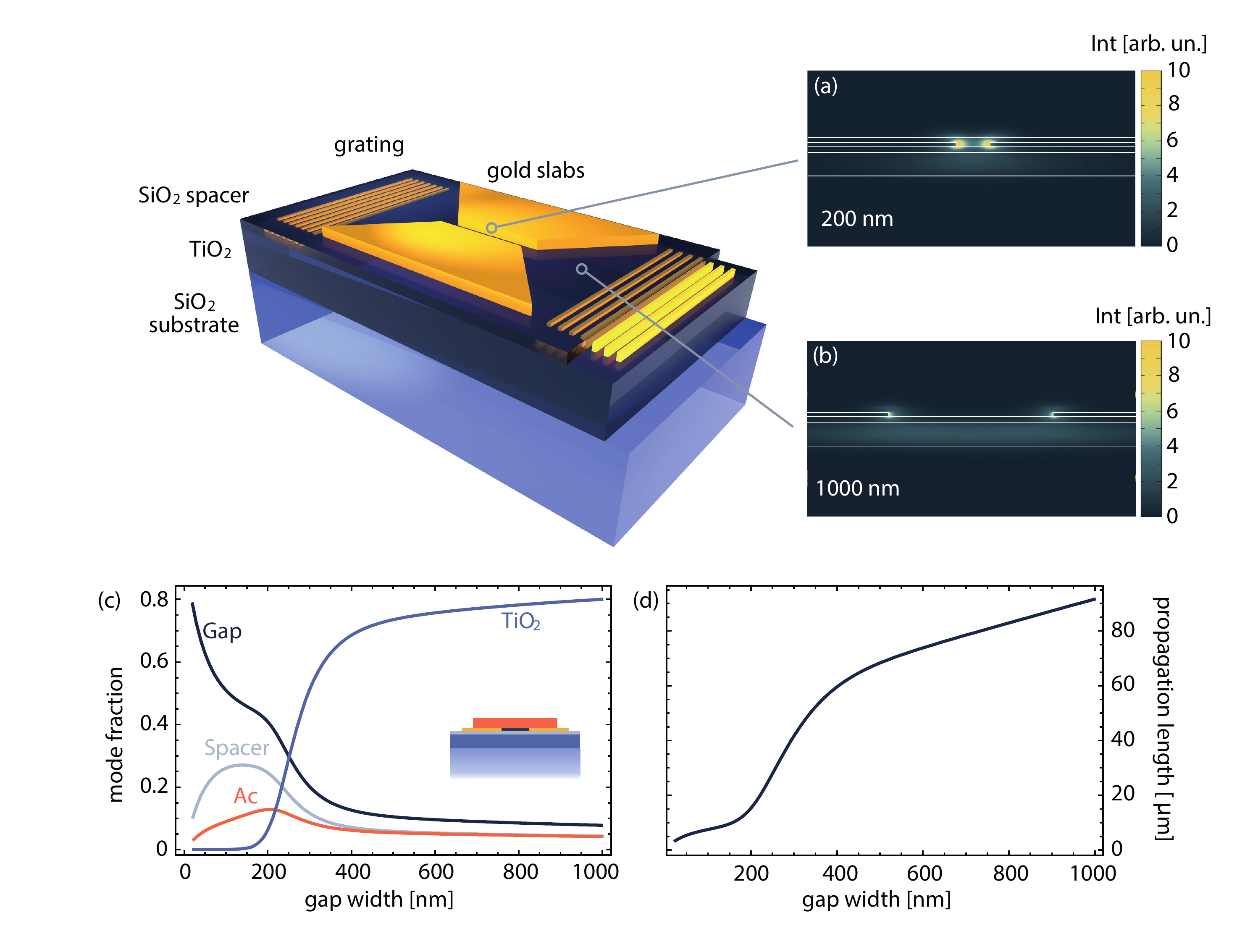}
	\caption{Schematic of the HGPW showing the layer structure. The two insets show the calculated intensity profiles of light in the HGPW for a gold gap of (a) 200~nm and (b) 1000~nm. (c) Distribution of the mode energy across the layers, as a function of channel gap width. The inset shows the various areas of the device considered. (d) Propagation length of the hybrid mode, again as a function of channel gap width.}
	\label{fig1}
\end{figure*}

Our HGPW is based on a design introduced by Lafone \textit{et al.}\cite{Lafone2014} and first fabricated and studied by Nielsen \textit{et al.}\cite{Nielsen2016,Nielsen2017}. The design was modified to operate at the $\sim$785~nm emission wavelength of DBT by replacing the silicon and gallium arsenide used in previously fabricated devices\cite{Nielsen2016,Nielsen2017,Nguyen2017} with titanium dioxide (TiO\textsubscript{2}). Figure~\ref{fig1} shows a schematic cross-section of the device, highlighting the underlying structure of a single HGPW. A 300~nm-thick layer of TiO\textsubscript{2} deposited on a silica (SiO\textsubscript{2}) substrate forms the base of our devices. Gold gratings deposited on the TiO\textsubscript{2} are covered by an 80~nm spacing layer of SiO\textsubscript{2}. Finally, two gold slabs are deposited on top of the SiO\textsubscript{2}. The gratings couple light in and out of the TiO\textsubscript{2}, where it propagates between the gold islands predominantly as a TE mode. As the gold islands taper to form a smaller gap the mode becomes increasingly hybridised with the plasmon mode on the edges of the gold. Figure~\ref{fig1}(a) and (b) show the distribution of electric field energy density calculated in COMSOL for channels of two different gap widths, and considering a layer of anthracene of $60$~nm placed over the structure. In Fig.~\ref{fig1}(a) the gold edges are separated by $1000$~nm and the energy is mostly in the TiO\textsubscript{2} layer. In Fig.~\ref{fig1}(b), where the gap is $200$~nm, most of the intensity is concentrated on the edges of the gold. There is no dielectric confinement of the mode in the lateral direction, as the TiO\textsubscript{2} layer covers the entire substrate and is shared by all the fabricated devices. However, the field is prevented from spreading in the plane by its coupling to the plasmon. This requires careful adjustment of the thickness of the SiO\textsubscript{2} spacer layer. By controlling the hybridisation in this way, it is possible to benefit from the field confinement, while still retaining a long enough propagation length that the field can emerge from the structure and be coupled out the other side. This tradeoff is shown in panels (c) and (d) of Fig.~\ref{fig1}, where the energy distribution across the device and the propagation length as a function of channel gap width are plotted. It is possible to see that for gap widths smaller than $300$~nm the field is extracted from the TiO\textsubscript{2} layer and is progressively concentrated in the gap region. This is accompanied by a decrease in propagation length, as the mode area is reduced.
%The calculated $1/e$ propagation length for the two widths presented here are 50~microns for the 1000~nm gap and 10~microns for the 200~nm gap -- much longer in both cases than the total length of the device.

\begin{figure}[t!]
	\centering
	\includegraphics[width=0.9\columnwidth]{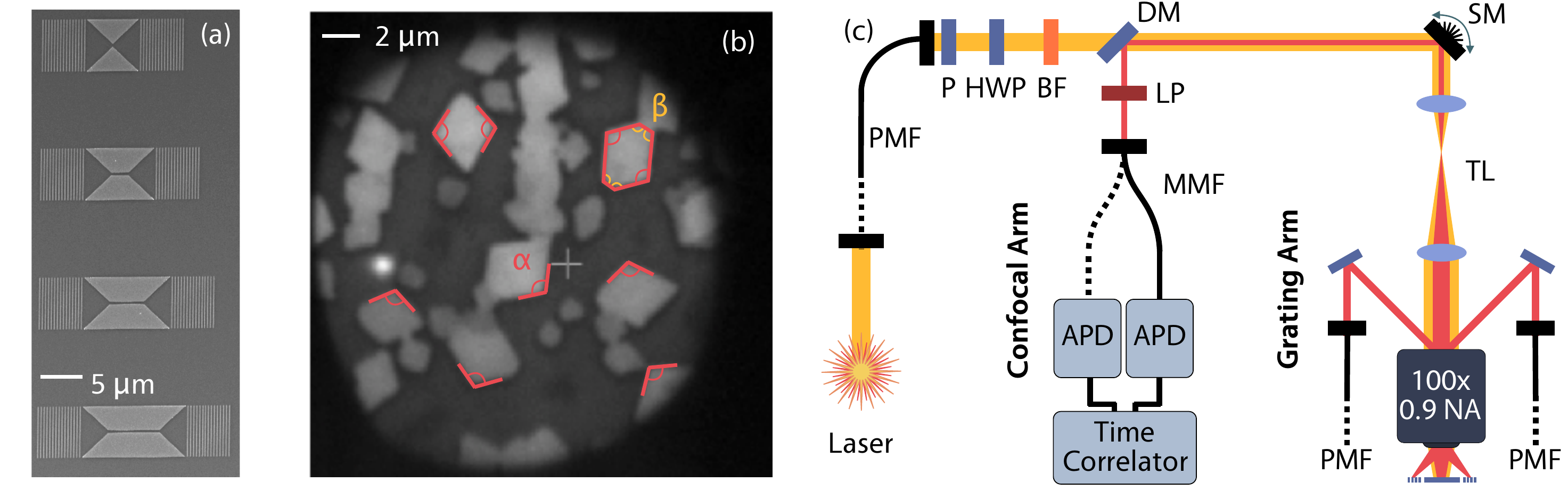}
	\caption{a) SEM image of four HGPWs whose plasmonic regions vary in length from 0 to 6 microns. b) White light image of anthracene crystals grown from supersaturated vapour. The angles $\alpha=109^\circ$ and $\beta=125^\circ$ are as expected for bulk crystals. c) Schematic of the confocal microscope used to study coupling of single molecules to hybrid plasmonic waveguides. PMF: polarisation-maintaining fibre. P: polariser. HWP: half-wave plate. BF: band-pass filter. DM: dichroic mirror. SM: steering mirrors. TL: telescopic lenses. LP: long-pass filter. MMF: multi-mode fibre. APD: avalanche photodiode.}
	\label{fig2}
\end{figure}

Figure~\ref{fig2}(a) shows a scanning electron microscope image of four HGPWs, where the plasmonic regions vary in length from 0 to 6 microns. Individual DBT molecules were deposited on top of our HGPW devices using a recently developed method\cite{Polisseni2016}. A 1~mMol solution of DBT in toluene was diluted in diethyl ether at a volume ratio of 1:2000, and spin-coated onto our sample. A glass vial was filled with 2~g of finely ground anthracene (Ac) powder and heated to 240$^\circ$C on a hotplate inside a glove box, purged of oxygen and filled with nitrogen gas to decrease the chance of the molecules photo-bleaching\cite{Kozankiewicz2014}. The vial was covered with a glass microscope cover slip. After roughly 2~minutes the cover slip was removed and the HGPW chip was put in its place. A growth time of about 2~minutes provided the desired coverage of anthracene crystals. This growth time results in crystals which are less than 100~nm thick. Figure~\ref{fig2}(b) shows a white light microscope image of the crystals, with the angles $\alpha$ and $\beta$ being those expected for anthracene\cite{Major2015}.
%%%%%%FIGURE

The confocal microscope used to study the waveguides is presented in Fig.~\ref{fig2}(c). A laser provided excitation light -- either a cw external cavity diode laser (Toptica) at $780$~nm wavelength, or a pulsed diode laser (PicoQuant) at $781$~nm. The laser light was cleaned in mode, polarisation and spectrum, then directed to a microscope objective (Nikon ApoPlan Fluo, 100x, 0.9 NA). A telecentric system of lenses and a set of two galvo mirrors allowed raster scanning of the focussed laser spot across the sample. The resulting fluorescence was separated from the pump light by a dichroic mirror (Semrock), and further filtered by two $800$~nm long-pass filters before being detected. A pellicle beam splitter was placed before the objective to add two additional beam paths, directing light to or from the gratings on the waveguide.

%%%%%%SECTION
\section{Results and discussion}
%%%%%%FIGURE
\begin{figure}[t!]
	\centering
	\includegraphics[width=0.5\columnwidth]{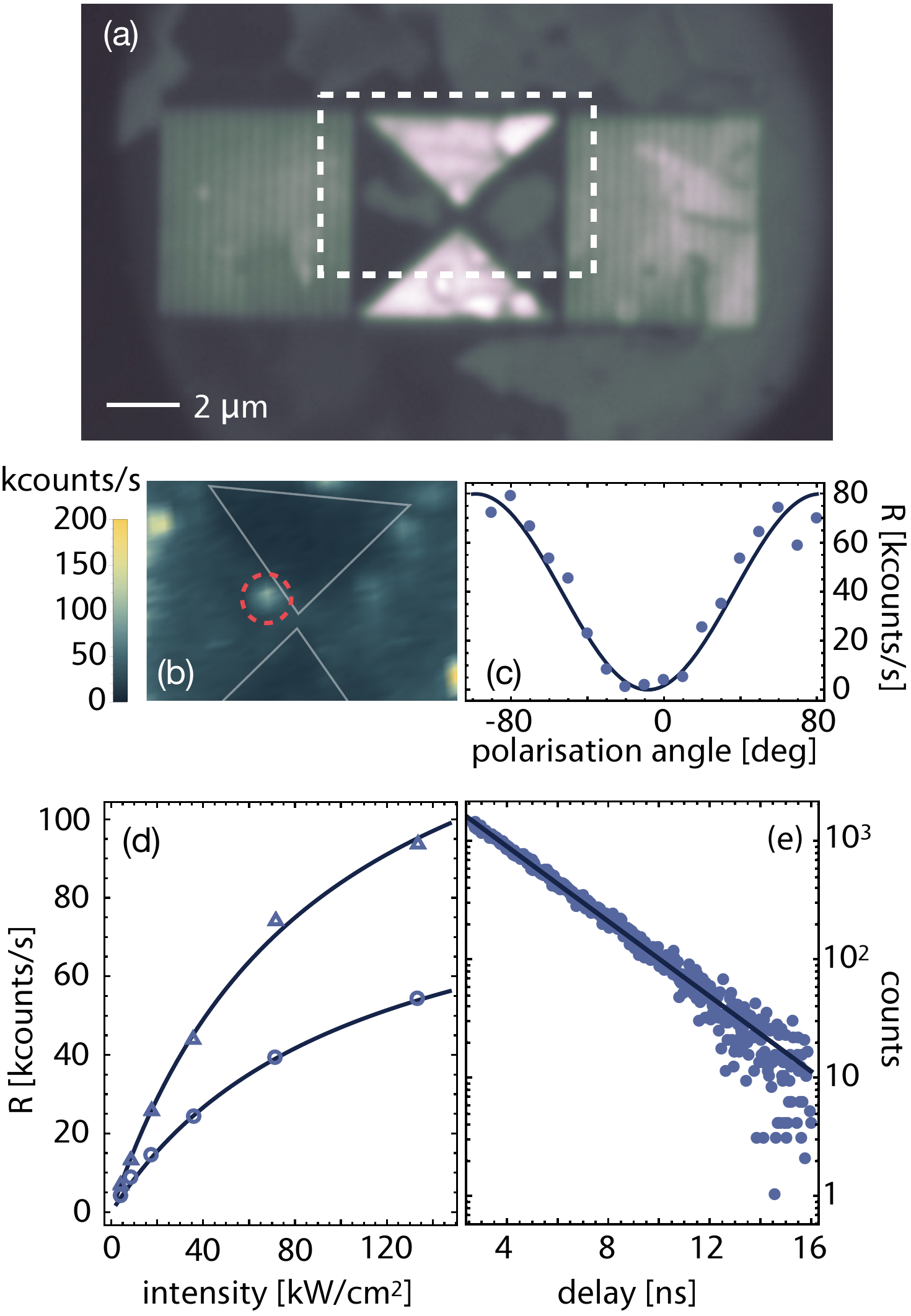}
	\caption{(a) White-light image of a HGPW showing input/output grating couplers and anthracene crystals on the surface. (b) Molecule fluorescence from the dashed box in (a). The gray lines outline the edges of the gold. (c) Collected count rate as a function of polarisation for the molecule outlined by the dashed red circle in (b). (d) Saturation curves for that molecule. Data points show count rates collected from the confocal microscope (triangles) and from a grating coupler (circles). The data is well fitted by the nonlinear function Eq.\eqref{eq:sat} (solid lines). (e) Pulsed laser measurement of the excited state lifetime of the same molecule.}
	\label{fig3}
\end{figure}
The HGPW of interest, shown in Fig.~\ref{fig3}(a), is 200~nm wide at the centre of the tapered region. We chose to work with this short HGPW as it has the lowest propagation loss. For coupling to longer HGPWs with higher loss, see Supplementary Materials. Figure~\ref{fig3}(b) shows fluorescence from the dashed box area, obtained by a confocal scan using cw light at 780~nm.  It should be noted that only one molecule seen in this confocal microscope scan resulted in photon detection events from a grating output. The signal from that DBT molecule, found near the centre of this device, is indicated by the red dashed circle. We first checked the orientation of this molecule by varying the polarisation of the excitation light while monitoring the detected photon rate, as reported in Fig.~\ref{fig3}(c). This showed that the optical dipole was only $6\pm2^\circ$ away from the optimum, that being perpendicular to the direction of propagation. We then varied the intensity of the excitation light while collecting fluorescence, both from the site of the molecule and from the grating to the left in Fig.~\ref{fig3}(a). In both collection arms the fluorescence rate saturates, as shown in Fig.~\ref{fig3}(d). The data points are well modelled by the saturation function
\begin{equation}
\label{eq:sat}
	R = R_{\infty} \frac{S}{1 + S}\, ,
\end{equation}
where $R_\infty$ is the asymptotic rate at high intensity and $S=I/I_\tx{sat}$, with $I$ being the peak intensity of the excitation light incident on the sample and  $I_\tx{sat}$ being the saturation intensity. These two fits gave saturation intensities $I^\tx{dir}_\tx{sat} = 90(8)~\tx{kW/cm}^2$ and $I^\tx{grat}_\tx{sat} =104(10)~\tx{kW/cm}^2$ for the direct and grating collection respectively, which are in good agreement with each other. The maximum photon rates were different, with values of $R^\tx{dir}_{\infty}=160(6)~\tx{kcounts/s}$ and $R^\tx{grat}_{\infty}=96(3)~\tx{kcounts/s}$, because of the differing collection efficiencies. The grating coupler on the right gave a count rate ten times lower. We do not think this was due to a fabrication imperfection because the throughput of this device was similar to that of the others on the same substrate, and we have simulated different molecule positions on the device and found no asymmetry in emission. The more likely explanation is that the surface patterning of the gold, or imperfections in the anthracene crystal, favoured emission into one direction over the other. The pulsed laser at 781~nm was then used to determine the decay time of the excited molecule. The semi-log plot in Fig.~\ref{fig3}(e) shows the measured probability distribution of delay times $t$ between excitation of the molecule and detection of a photon, after correcting for the background count rate. A fit using the function $A\, e^{-t/\tau}$ gave the lifetime of the excited state of the molecule as $\tau = 2.74(2)~\tx{ns}$. This is slightly shorter than the expected 3--6~ns for DBT in anthracene\cite{Nicolet2007a,Trebbia2009,Toninelli2010,Major2015,Polisseni2016,Grandi2016}. This could be because the decay rate of the molecule is enhanced by its coupling to the waveguide (see Supplementary Materials), or possibly that the non-radiative decay rate was increased.

\begin{figure}[t!]
	\centering
	\includegraphics[width=0.5\columnwidth]{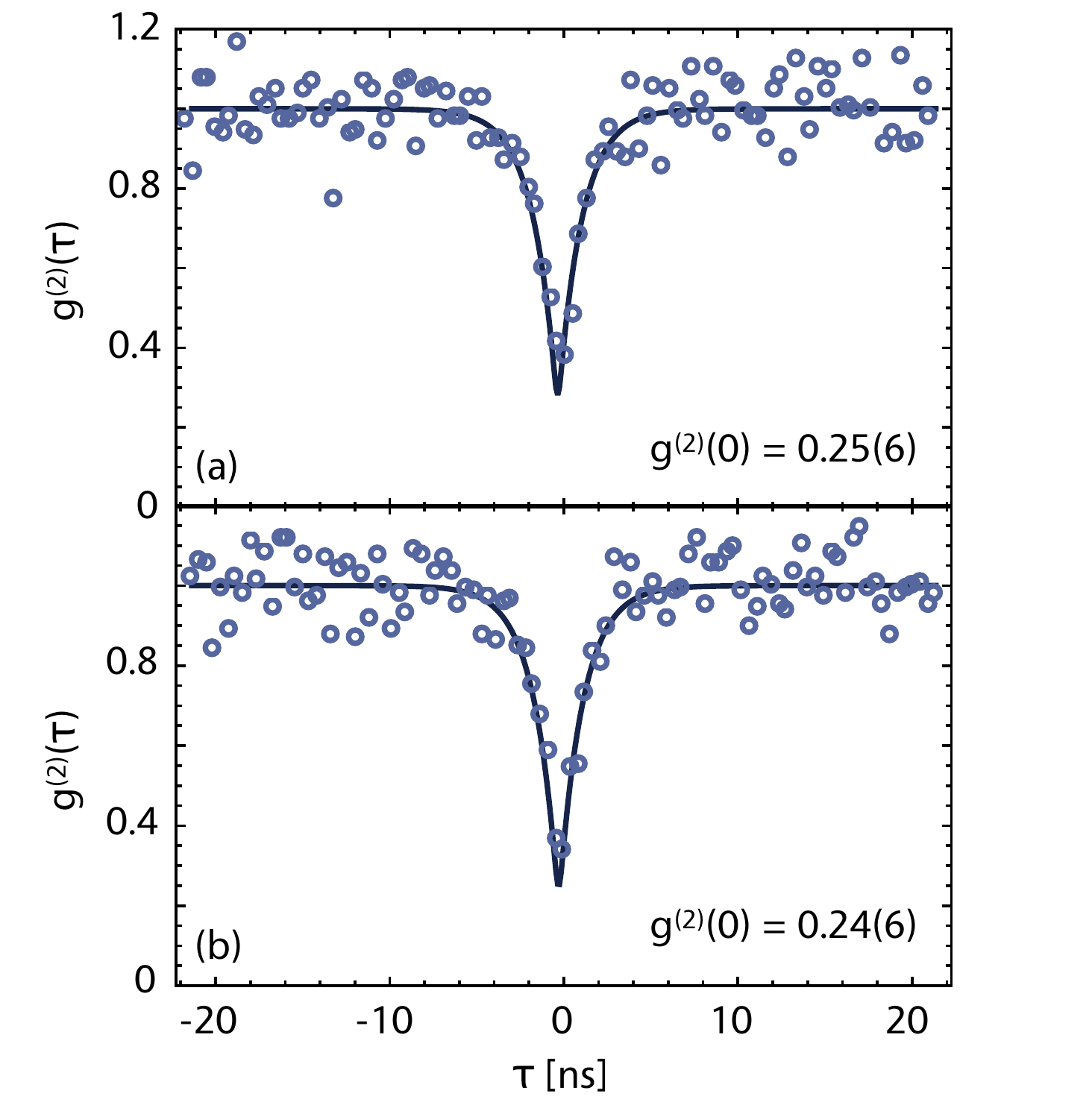}
	\caption{Correlation functions $g^{(2)}(\tau)$ for the DBT fluorescence. (a) Auto-correlation of light collected directly from the molecule through the confocal microscope after a beamsplitter. (b) Cross-correlation of light collected from the left grating and from the confocal microscope. Dots show data. Solid curves are fits using Eq.~\ref{eq:g2}. In both cases $g^{(2)}(0)$ is well below 0.5 showing clear anti-bunching.}
	\label{fig4}
\end{figure}

To confirm that this was indeed a single molecule we measured the second-order correlation function $g^{(2)}(\tau)$ for the emitted light, while exciting the molecule at a saturation level $S\approx 1$. As a single molecule can only emit one photon at a time, we expect that $g^{(2)}(0)=0$ in the ideal case\cite{Loudon2000}. To determine $g^{(2)}(\tau)$ for the fluorescence collected directly from the molecule, a 50:50 multi-mode fibre splitter (Fig. 2(c)) divided the light between two avalanche photodiodes, and a time correlating card recorded the histogram of start--stop intervals in the standard way. The upper panel of Fig.~\ref{fig4} shows the data points, together with a fit to the function\cite{Grandi2016}
\begin{equation}
\label{eq:g2}
	\gtwo = 1 - B \, e^{-(1+S)\frac{t}{\tau}}
\end{equation}
where $B=1-\gtwoZ$ was the only free parameter. This $g^{(2)}(\tau)$ exhibits clear anti-bunching, with the fit giving $g^{(2)}(0)=0.25(6)$. Next, we removed the fibre splitter and instead measured the time correlation between the light from the molecule and that collected from the grating. This gave the $g^{(2)}(\tau)$ in the lower panel of Fig.~\ref{fig4}, where $g^{(2)}(0)=0.24(6)$. The two values of $g^{(2)}(0)$ agree. The fact that $\gtwoZ<0.5$ without any correction (e.g. for background counts, dark counts or possible nearby emitters), signifies that we were indeed collecting fluorescence predominantly from a single molecule in both cases. After convolving Eq.~\eqref{eq:g2} with the Gaussian instrument response function due to detector timing jitter (standard deviation of 455~ps), both data sets gave $g^{(2)}(0)=0.20(2)$, a value that is consistent with the signal-to-background ratio found in each case.

%%%%%%FIGURE

%%%%%%SECTION
\section{Deducing the coupling efficiency}
We use the detected fluorescence rate to estimate the coupling factor $\beta$, defined as the fraction of photons coupled into the waveguide:
\begin{equation}
	\beta = \frac{\Gamma_\tx{wg}}{\Gamma_\tx{tot}},
\end{equation}
where $\Gamma_\tx{wg}$ is the rate at which the molecule emits photons into the waveguide and $\Gamma_\tx{tot}$ is its total emission rate. The rate at which photons are detected from the grating is $R_\tx{grat}=\Gamma_\tx{wg}\eta_\tx{grat}$, where $\eta_\tx{grat}$ is the efficiency for coupling light out from the waveguide, collecting it, and detecting it using the APD. We can further write $R_\tx{grat}=R^\tx{grat}_\infty S/(1+S)$, where $R^\tx{grat}_\infty$ is the fully saturated rate detected from the grating.  Similarly, the total emission rate can be written as $\Gamma_\tx{tot}=\alpha \, \Gamma_1S/(1+S)$, where $\Gamma_1=1/\tau$. In the limit of large $S$, this tends to $\alpha\Gamma_1$, where $\alpha$ lies between $0.5$ and $1$, depending on the excitation scheme\cite{Schofield2018}. Thus the coupling factor $\beta$ is given by
\begin{equation}
\label{eq:beta}
	\beta = \frac{\tau}{\alpha} \frac{R^\tx{grat}_\infty}{\eta_\tx{grat}}.
\end{equation}
We have measured $\tau$ and  $R^\tx{grat}_\infty$ (see above), and for room-temperature DBT excited at $780$~nm, we know\cite{Schofield2018} that $\alpha=0.555(10)$. That leaves us needing to assess $\eta_\tx{grat}$.

At room temperature the DBT molecule emits photons over a $\sim20$~THz wide frequency range\cite{Schofield2018}. This is broad enough that $\eta_\tx{grat}$ has to be determined by convolving the emission spectrum with the frequency-dependent outcoupling/collection/detection efficiency. In separate experiments (see Supplementary Materials), we have measured the frequency-dependent output coupling from the waveguide through the grating. With a peak value of $10\%$ at $800\,$nm, dropping to $2\%$ at $765\,$nm and $830\,$nm, this is the main loss-factor contributing to $\eta_\tx{grat}$. We have also accounted for the frequency-dependent transmission of all the other optical elements (see Supplementary Materials). The result is $\eta_\tx{wg}= 4.1(5)\times 10^{-3}$, which leads to a result for the coupling efficiency of the molecule to the waveguide of $\beta = 11.6(1.5)\%$. This result agrees well with finite difference time domain simulations; while these didn't show the same asymmetry between left and right grating emission, they resulted in a total coupling efficiency to the waveguide mode of $11.5\%$ (see Supplementary Materials). Other experiments have observed similarly high coupling to dielectric waveguides\cite{Turschmann2017,Faez2014,Lombardi2017}, but the use of plasmonics in our case opens the possibility of much stronger coupling in the future.

%%%%%%SECTION
\section{Summary and future prospects}
 We have observed the coupling of a single DBT molecule to a HGPW made from a multi-layer dielectric slab patterned with gold structures. Measurements on the molecule itself gave values for the in-plane orientation, excited state lifetime and saturation intensity, and confirmed through the second-order correlation function $g^{(2)}(\tau)$ of the emitted light that this was a single molecule. We also detected the light coupled out of the waveguide by a grating, and measured the cross-correlation of this light with that observed at the molecule. This showed that the light at the grating was indeed emitted by the molecule. The photon count rate detected at the grating was used to infer the efficiency $\beta$ with which the molecule radiated photons into the guide. These measurements were made at room temperature, where phonon-induced dephasing of the optical dipole\cite{Grandi2016, Schofield2018} makes the photons spectrally broad.  Such photons can be useful for communication and imaging, but not for applications that require quantum interference such as linear optical quantum computing or quantum simulation. In the future we will look for molecules coupled to HGPW at liquid helium temperatures, where decoherence should be minimized so the spectrum should exhibit a Fourier-limited spectral width.

These measurements were made on a single device with a gold gap width of 200~nm, which is not expected to give a large enhancement of the photon emission rate. We plan to look for molecules coupled to guides with smaller gap sizes where the coupling should be strongly enhanced. One of the main limitations in our device was the low contrast of refractive index between the titanium dioxide and the silica substrate.  To improve on this, we have simulated the case of gallium phosphide (GaP) on silica, as it presents a refractive index comparable to silicon and shows low losses in the near infrared\cite{Schnieder2018}. Modifying our structure\cite{Nielsen2016} with a thinner spacing layer to retain mode hybridisation, we have found that a total coupling efficiency higher than $50\%$ can be achieved by placing a DBT molecule in a $500$~nm long waveguide with a gap width of $100$~nm.

Finally, in the future it should be possible to use similar plasmonic structures to shift the waveguide mode adiabatically in and out of low-loss dielectric ridge waveguides, creating regions of strong light-matter interaction at precise locations in a low-loss integrated photonic network.

%%%%%%SECTION
\begin{acknowledgments}
We thank Jon Dyne, Giovanni Marinaro, and Valerijus Gerulis for their expert mechanical and electrical workshop support. This work was supported by EPSRC (EP/P030130/1, EP/P01058X/1, EP/M013812/1, EP/P030017/1, and EP/G037043/1), The Leverhulme Trust (RPG-2016-064), dstl (DSTLX1000092512), The Royal Society (RP110002, UF160475), a Natural Sciences and Engineering Research Council of Canada (NSERC) scholarship, the European Commission (Marie Curie Action - Initial Training Networks: Frontiers in Quantum Technology Project 317232), and a Marie Sk\l odowska Curie Individual Fellowship (Q-MoPS, 661191).
\end{acknowledgments}

%%%%%%SECTION

\bibliography{library}

\end{document}